\documentclass[sigconf]{acmart}
\usepackage{algorithm}
\usepackage{algorithmic}
\usepackage{multirow}
\usepackage[section]{placeins}

\AtBeginDocument{%
  \providecommand\BibTeX{{%
    \normalfont B\kern-0.5em{\scshape i\kern-0.25em b}\kern-0.8em\TeX}}}

\setcopyright{acmcopyright}
\copyrightyear{2018}
\acmYear{2018}
\acmDOI{10.1145/1122445.1122456}

\acmConference[Woodstock '18]{Woodstock '18: ACM Symposium on Neural
  Gaze Detection}{June 03--05, 2018}{Woodstock, NY}
\acmBooktitle{Woodstock '18: ACM Symposium on Neural Gaze Detection,
  June 03--05, 2018, Woodstock, NY}
\acmPrice{15.00}
\acmISBN{978-1-4503-XXXX-X/18/06}

\begin{document}

\title{GIMIRec: Global Interaction Information Aware Multi-Interest Framework for Sequential Recommendation}

\author{Jie Zhang}
\email{1020041203@njupt.edu.cn}
\affiliation{%
	\institution{Nanjing University of Posts and Telecommunications}
	\streetaddress{9$^{th}$ Wenyuan Road}
	\city{Nanjing}
	\state{Jiangsu}
	\country{China}
	\postcode{210046}
}

\author{Ke-Jia Chen}
\email{chenkj@njupt.edu.cn}
\affiliation{%
	\institution{Nanjing University of Posts and Telecommunications}
	\streetaddress{9$^{th}$ Wenyuan Road}
	\city{Nanjing}
	\state{Jiangsu}
	\country{China}
	\postcode{210046}
}

\author{Jingqiang Chen}
\email{cjq@njupt.edu.cn}
\affiliation{%
	\institution{Nanjing University of Posts and Telecommunications}
	\streetaddress{9$^{th}$ Wenyuan Road}
	\city{Nanjing}
	\state{Jiangsu}
	\country{China}
	\postcode{210046}
}

\renewcommand{\shortauthors}{Zhang and Chen, et al.}

\begin{abstract}
Sequential recommendation based on multi-interest framework models the user's recent interaction sequence into multiple different interest vectors, since a single low-dimensional vector cannot fully represent the diversity of user interests. However, most existing models only intercept users' recent interaction behaviors as training data, discarding a large amount of historical interaction sequences. This may raise two issues. On the one hand, data reflecting multiple interests of users is missing; on the other hand, the co-occurrence between items in historical user-item interactions is not fully explored. To tackle the two issues, this paper proposes a novel sequential recommendation model called `` {\bfseries G}lobal {\bfseries I}nteraction Aware {\bfseries M}ulti-{\bfseries I}nterest Framework for Sequential {\bfseries Rec}ommendation (GIMIRec)''. Specifically, a global context extraction module is firstly proposed without introducing any external information, which calculates a weighted co-occurrence matrix based on the constrained co-occurrence number of each item pair and their time interval from the \textit{historical} interaction sequences of all users and then obtains the global context embedding of each item by using a simplified graph convolution. Secondly, the time interval of each item pair in the \textit{recent} interaction sequence of each user is captured and combined with the global context item embedding to get the personalized item embedding. Finally, a self-attention based multi-interest framework is applied to learn the diverse interests of users for sequential recommendation. Extensive experiments on the three real-world datasets of Amazon-Books, Taobao-Buy and Amazon-Hybrid show that the performance of GIMIRec on the Recall, NDCG and Hit Rate indicators is significantly superior to that of the state-of-the-art methods. Moreover, the proposed global context extraction module can be easily transplanted to most sequential recommendation models.
\end{abstract}




\maketitle

\section{Introduction}
The recommendation system predicts users' preferences for items based on their historical behaviors. For example in e-commerce application, the users are often recommended some commodities based on their past purchase behaviors. Deep neural networks, especially graph neural networks (GNNs) \cite{GNN, GNN1}, have now been applied on the recommendation systems, such as GRU4Rec \cite{GRU4Rec}, GGNN \cite{GGNN} and SR-GNN \cite{SRGNN}, achieving better performance than traditional collaborative filtering methods \cite{CF1,CF2}. However, the graph-based methods are often with high complexity as the result of the large scale of users and items in real world \cite{LightGCN}. The sequential recommendation has recently drawn more research attention as it starts with each user's own interaction data, models the users’ preferences with only recent interaction behaviors and then gives personalized recommendation for each user.

The main effort of sequential recommendation is to learn the representations of users and items. Since the interests of users are usually diverse and difficult to be represented by a single low-di-mensional vector, the sequential recommendation methods based on multi-interest framework have emerged \cite{MIND, ComiRec, PIMI}. In these methods, the number of representation vectors for each user is increased from 1 to $\mathcal{K}$, where each vector represents one interest of the user. Multiple interests can be captured by an improved capsule network \cite{MIND} or a self-attention mechanism \cite{ComiRec} and ultimately a more personalized recommendation based on $\mathcal{K}$ interests is provided. 

Most of the existing multi-interest frameworks for sequential recommendation use only users’ recent interaction sequence as it is believed that the user’s next interaction item is more closely related to his/her recent behaviors. However, we believe that the complete historical interactions of all users (referred to as the \textit{global context} in this paper) can be better leveraged to reflect the diversity of users’ interests and potential relations between items.

An example shown in Figure 1 is to illustrate how the global context affects the prediction of the 10\textit{th} interactive item for each user although this item did not appear in each user's recent interactions. Since the \textit{popcorn} and \textit{wine} interaction sequence can be found in user A’s early historical sequence, the 10\textit{th} interactive item for user B will be probably the \textit{wine}. The \textit{cake} could be recommended for user A instead of \textit{battery} due to the co-occurrence of \textit{game console} and \textit{cake} in user C's historical sequence. The example indicates that the user’s behaviors could be predicted with the reference of other users’ historical behaviors though different users may have different preferences.

\begin{figure}[h]
	\centering
	\includegraphics[width=\linewidth]{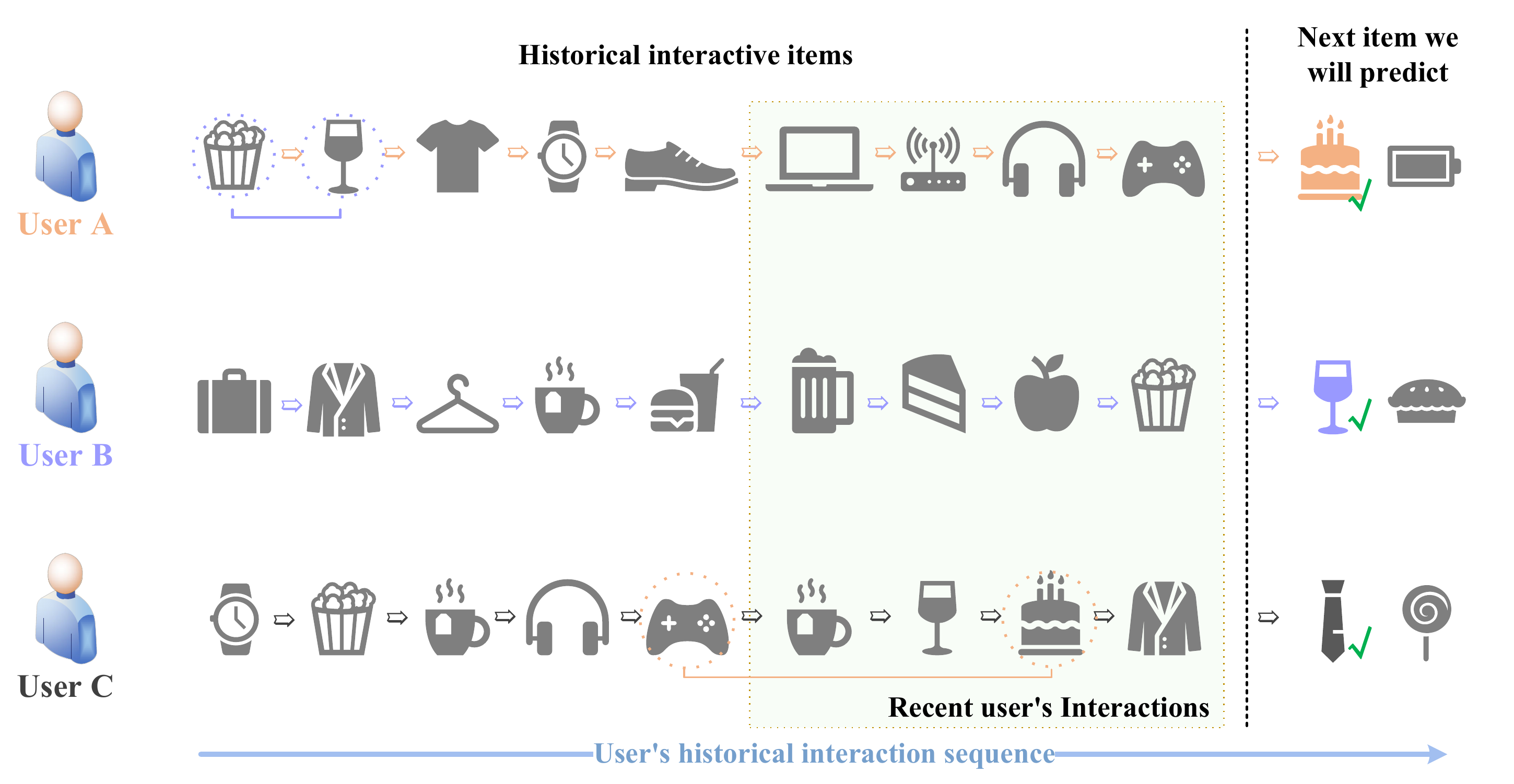}
	\caption{An example of users' interaction sequences.}
	\Description{}
\end{figure}

To our knowledge, these have been some sequential recommendation models exploring users' complete sequence to learn their preferences \cite{1DMAN, 3AAAI, 1SIGIR, 1WWW, 2AAAI}. But they focus more on each user's personalized behaviors but do not take the global context into consideration. 

Therefore, this paper proposes a global interaction information aware multi-interest framework for sequential recommendation (named GIMIRec) with the main contributions as below:

\begin{itemize}
	\item A new multi-interest sequential recommendation model is proposed, which is equipped with a global context extraction module to fully utilize the historical sequences of all users. This module can also provide a performance boost for the recommendation models based on representation learning.
	
	\item A lightweight approach is designed to calculate the global context embedding of each item, based on the weighted co-occurrence matrix fusing the constrained co-occurrence number of item pairs and their time intervals in the global context. 
	
	\item Our model achieves the state-of-the-art performance on three real-world datasets without introducing any other supportive information (such as item attributes or user’s social information) other than interactions and timestamps. 
\end{itemize}

\section{Related Work}
{\bfseries Sequential recommendation.} Traditional recommendation methods, such as matrix decomposition and collaborative filtering \cite{CF1, CF2} , represent interaction behaviors as a user-item matrix and explore their potential relation. With the development of deep neural networks, different recommendation methods combined with deep learning have emerged, such as the neural collaborative filtering method NGCF \cite{NGCF}. However, it is hard for the traditional recommendation system to manage huge but sparse interaction data. To solve this problem, sequential recommendation methods are proposed. The objective of sequential recommendation is to learn each user's preferences from the historical interaction behaviors, and then to achieve recommendation. Typical methods include GRU4Rec \cite{GRU4Rec} and SASRec \cite{SASRec} where the recurrent neural network and the attention mechanism are used respectively. Later, the time of the interaction occurrence is also taken into account. For example, a parallel time-aware mask network is proposed in MTIN \cite{MTIN}, which extracts a variety of time information through multiple networks to enhance the short video recommendation effect. Ti-SASRec \cite{TiSASRec} is another method incorporating time interval of every two items in sequences to further improve the performance of SASRec \cite{SASRec}.

{\bfseries Multi-interest framework for sequential recommendation.} User preferences may contain a variety of interests. The existing sequential recommendation systems eventually model users’ preferences as a single low-dimensional vector while the users may have more than one single interest. MIND \cite{MIND} was the first to model each user with $\mathcal{K}$ vectors, where different vectors represent different interests. The extraction of multi-interests is similar to the clustering process and the behavior-to-interest (B2I) approach is proposed to obtain $\mathcal{K}$ interests of users through a capsule network. Subsequently, \textit{Cen} et al. \cite{ComiRec} proposed a multi-interest extraction method based on self-attention mechanism (ComiRec-SA) to instead of capsule network. The latest method PIMI \cite{PIMI} introduced the time and periodic information in users’ recent interaction sequences into the model. The self-attention mechanism was used to capture users’ multi-interests and achieved the state-of-art performance. However, all the above multi-interest based sequential recommendation models, including PIMI, do not fully utilize the historical interactions of all users, thus lacking the global context of items.

{\bfseries Graph neural networks.} In recommendation systems, users’ historical interaction behaviors can be converted into a graph structure, where users and items are represented as nodes and their interactions are represented as edges. Graph neural networks (GNNs) \cite{GNN1, GNN} have been proved to better learn the representations of users and items. In non-sequential recommendation models, such as NGCF \cite{NGCF}, LightGCN \cite{LightGCN} and DGCF \cite{DGCF}, simplified GNNs are used on the bipartite graph composed of users and items to generate their embeddings. In sequential recommendation models such as SR-GNN \cite{SRGNN},  GNNs are also widely used to represent user’s sequence and to capture user's preferences.

\section{Problem Definition}
Before describing the proposed model in detail, the formal definitions of sequential recommendation and multi-interest framework are given as follows:

{\bfseries Definition 1: Sequential Recommendation.} Assuming that $\mathcal{U}$ represents a collection of all users and $\mathcal{I}$ represents a collection of all items. For a given user $u\in\mathcal{U}$, his/her interaction sequence $\mathcal{S}_u$ is defined with an item sequence  $\mathcal{I}_u = \left\{i_1^u, i_2^u, \cdots ,i_{\mathcal{N}}^u\right\}$ and a timestamp sequence of each interaction occurrence $\mathcal{T}_u = \left\{t_1^u, t_2^u, \cdots ,t_{\mathcal{N}}^u\right\}$.  The task of sequential recommendation is to predict the next interactive item $i_{\mathcal{N}+1}^u$ for the user $u$ based on $\mathcal{S}_u$. 

{\bfseries Definition 2: Multi-Interest Framework.} It denotes a framework designed for sequential recommendation. According to Definition 1, each user $u$ will be represented as a $d$-dimensional embedding by learning from his/her interaction sequence. However, one single vector representation may not be sufficient to describe the user’s multi-interest preferences. In the multi-interest framework, $\mathcal{K}$ $d$-dimensional embeddings of each user will be learned for sequential recommendation.

\section{Method}
\subsection{Overview}
The main purpose of our model GIMIRec is to better capture the multiple interests of each user from the global context information, which is calculated from the historical sequences of all users. The overall model consists of the following modules.

{\bfseries Global Context Extraction Module.} The module is basically to collect the co-occurrence number and time interval for any item pair that appear one after another in each user’s historical interaction sequence for the calculation of the weighted co-occurrence matrix. Subsequently, a simplified graph convolution is used on the above matrix to obtain the \textit{global context embedding} of each item.

{\bfseries Recent Time Interval Representation Module.} Though we believe the global context will better reflect the diversity of user preferences, the user's recent interactions directly reflect user’s recent preferences. So this module first identifies the recent interaction sequence based on the predetermined length $L_{rec}$, and then encodes the time interval between any item pair within the sequence to obtain the \textit{time interval embedding} for recent items.

{\bfseries Aggregation Module.} For each item in the user’s recent interaction sequence, the global context embedding and the time interval embedding are merged in this module to get the hybrid embedding. For each user, a graph neural network is used to learn the final \textit{personalized embedding} by aggregating the above hybrid embeddings in his/her recent interaction sequence.

{\bfseries Multi-Interest Extraction Module.} The module extracts $\mathcal{K}$ interest vectors from the personalized embedding of user’s recent sequence by the self-attention mechanism \cite{MTHEADATT}. For model training, one interest vector that can best represent user’s current preference is selected from $\mathcal{K}$ interest vectors.

Finally, the trained model is used to predict the user’s next $N$ interactive items (i.e., Top-$N$ recommendation) in the test set according to user’s $\mathcal{K}$ interests.

The architecture of GIMIRec is demonstrated in Figure 2, and each module will be detailed in the following subsections.

\begin{figure}
	\centering
	\includegraphics[width=\linewidth]{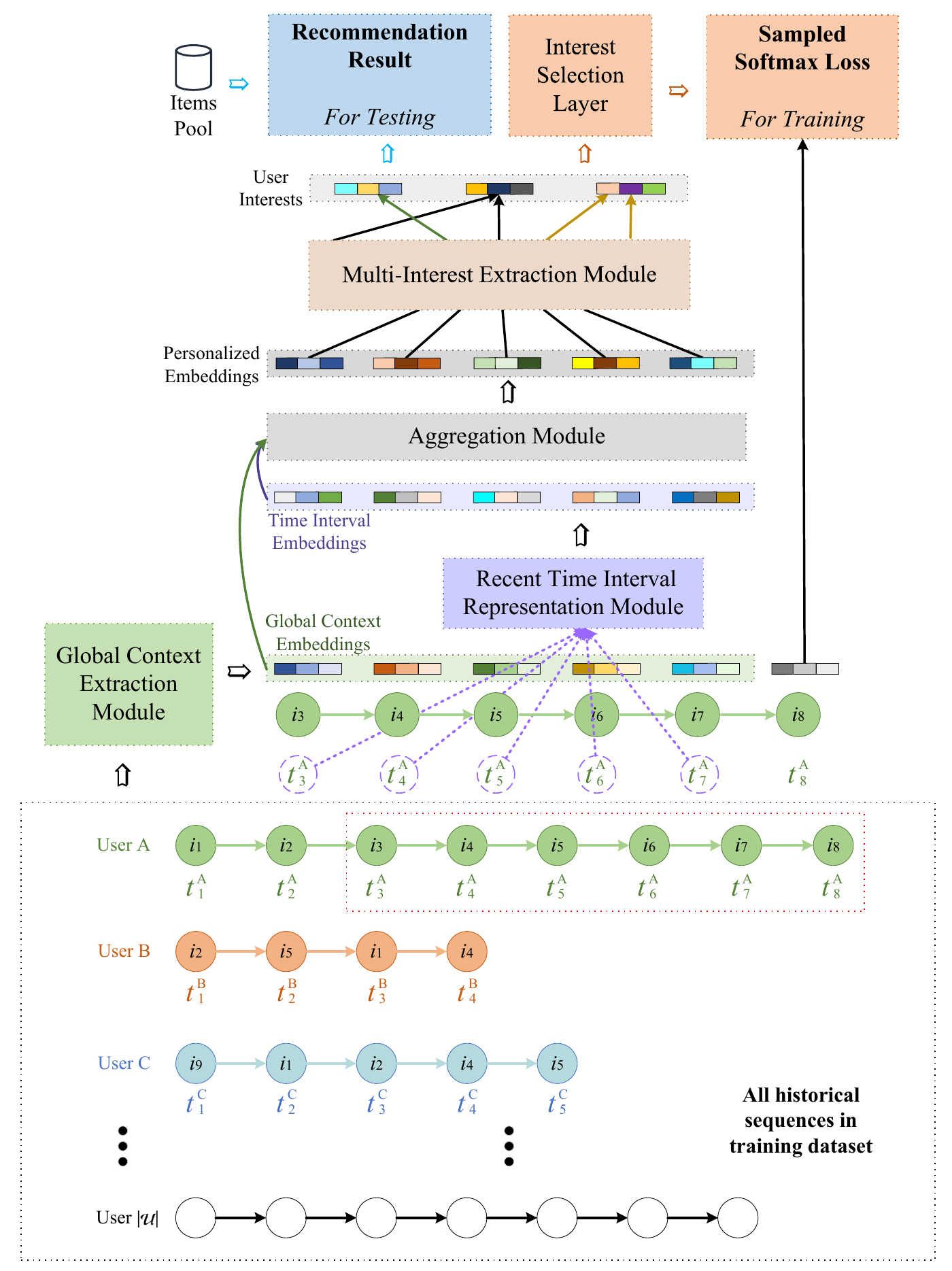}
	\caption{The architecture of the proposed model GIMIRec. }
	\Description{An overview of our model.}
\end{figure}

\subsection{Global Context Extraction Module}
The existing multi-interest frameworks for sequential recommendation extract the user’s interests only from $L_{rec}$ items with which user recently interacted. Historical interactions earlier than $L_{rec}$ items are discarded. 

The Global Context Extraction (GCE) Module tries to learn the global context embedding of items from all historical user interactions. Figure 3 illustrates how the GCE module processes historical interaction sequences with timestamps to get the weighted co-occurrence matrix $\mathcal{A}'$ by taking three users A, B and C as example. On the right-hand side of the Figure 3, a simplified graph convolution is used on $\mathcal{A}'$ to obtain the global context embedding of each item.

Here, the weighted co-occurrence matrix $\mathcal{A'}$ is defined as a combination of $k$-hop ($k=1, 2, 3$) item co-occurrence matrices. For each $\mathcal{S}_u$, the $k$-hop item pair is defined as a directed <$i_n^u$, $i_{n+k}^u$> ($n\in\mathcal{N}$). For example, user B in Figure 2 has four interacted items, so the 1-hop item pairs are <$i_2$, $i_5$>, <$i_5$, $i_1$> and <$i_1$, $i_4$>, the 2-hop item pairs are <$i_2$, $i_1$>, <$i_5$, $i_4$> and the 3-hop item pair is <$i_2$, $i_4$>.

The detailed steps of the GCE module are as follows.

Firstly, the $k$-hop item pairs are filtered out from the historical interaction sequences $\mathcal{I}_u = \left\{i_1^u, i_2^u, \cdots ,i_{\mathcal{N}}^u\right\}$ and the corresponding timestamp sequence $\mathcal{T}_u = \left\{t_1^u,t_2^u, \cdots ,t_{\mathcal{N}}^u\right\}$, $u\in\mathcal{U}$. In order to better capture the relation between items, a time interval threshold $L_{time}$ is set. The two items in $\mathcal{S}_u$ with the time interval less than $L_{time}$ can become a co-occurrence item pair. Given the large number of items in the sequences, we only consider 1-hop, 2-hop and 3-hop item pairs to reduce the amount of calculation. For convenience, triplet collections $\mathcal{R}_k (k=1, 2, 3)$ (Eq. 1) are used to save the id of two items and their time interval for three types of hop. Since the user id is no longer functional here, the superscript $u$ is replaced with $j$, representing one co-occurrence of the item pair <$i_\mu, i_\upsilon$>. For different $j$, the time interval of <$i_\mu, i_\upsilon$> could be different.

\begin{equation}
	\mathcal{R}_k = \left\{(<i_\mu^j, i_\upsilon^j, t_\upsilon^j - t_\mu^j>)  |  t_\upsilon^j - t_\mu^j \leq L_{time} \right\}
\end{equation}

As illustrated in Figure 2, there are three users interacted with 9 items in total, thus initially generating 14 $1$-hop item pairs, 11 $2$-hop item pairs and 8 $3$-hop item pairs. Assuming that the time intervals $t_4^A-t_3^A$, $t_6^A-t_5^A$, $t_7^A-t_6^A$, $t_3^B-t_2^B$ and $t_2^C-t_1^C$ exceed the threshold $L_{time}$, the final number of item pairs would be 9, 3 and 1 for $1$-hop, $2$-hop and $3$-hop respectively shown in Figure 3.

\begin{figure}
	\centering
	\includegraphics[width=\linewidth]{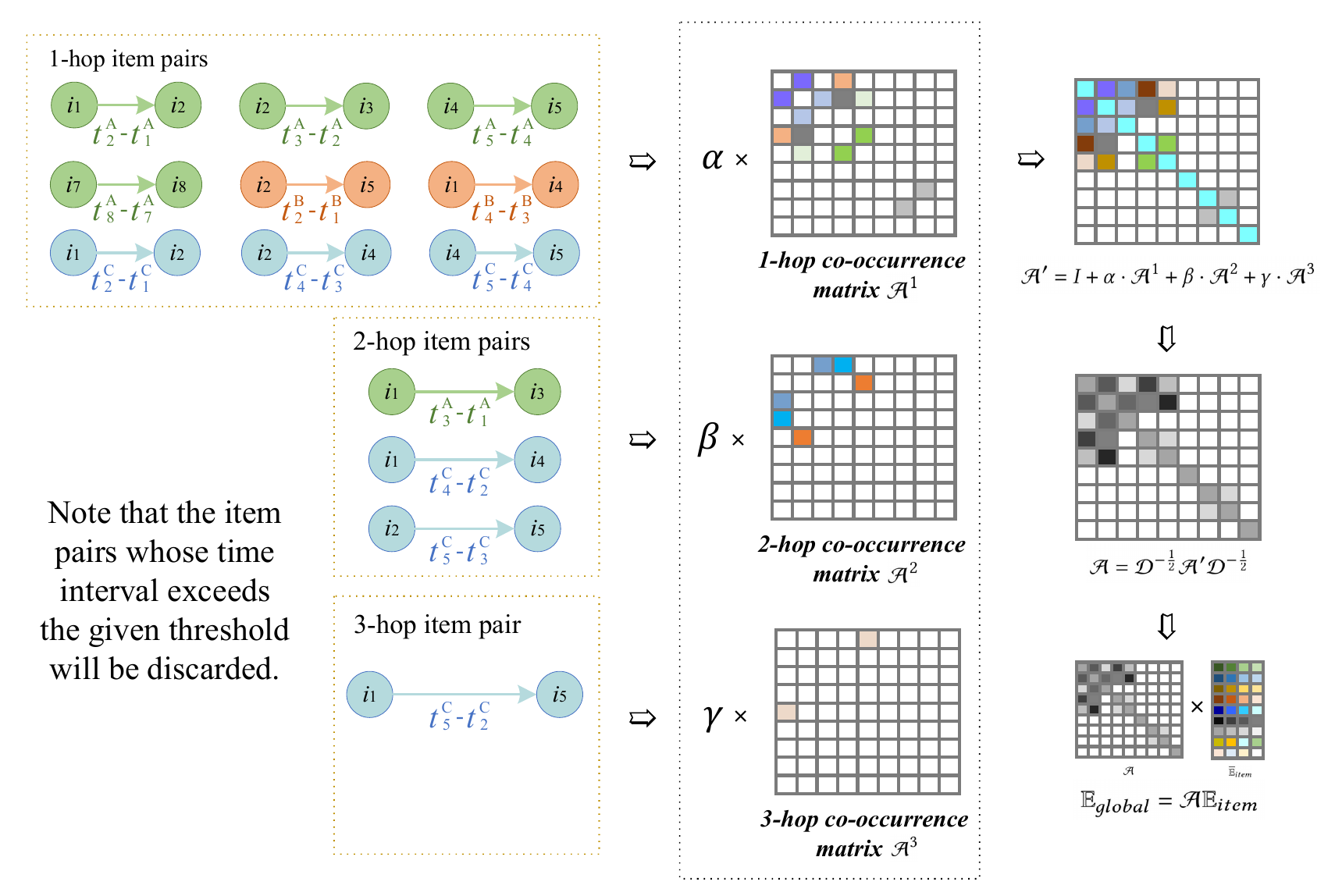}
	\caption{An example of Global Context Extraction Module.}
	\Description{An example of Global Context Extraction Module.}
\end{figure}

Secondly, three co-occurrence matrices $\mathcal{A}^k (k=1,2,3)$ are calculated. Intuitively, the item pair will have closer relation if they appear more frequently. Therefore, the entry in $\mathcal{A}^k$ will be higher with shorter time intervals and higher frequency. For each collection $\mathcal{R}_k (k=1, 2, 3)$ defined above, the time interval of an item pair <$i_\mu, i_\upsilon$> is first calculated by Eq. 2

\begin{equation}
	p_{<\mu, \upsilon>_k}^j = a \cdot {\frac{L_{time} - (t_\upsilon^j - t_\mu^j)}{L_{time}} } + b ,
\end{equation}
where $a$ and $b$ are hyperparameters satisfying $a+b=1$ and the time interval of <$i_\mu, i_\upsilon$> is normalized to [0, 1]. 

Eq. 3 sums up all the time intervals of <$i_\mu, i_\upsilon$>. 

\begin{equation}
	q_{\mu\upsilon}^k = \sum {p_{<\mu, \upsilon>_k}^j}
\end{equation}

Eq. 4 symmetrizes each matrix to facilitate the subsequent graph convolution, which usually performs better on an undirected graph \cite{GCN}. 

\begin{equation}
	\mathcal{A}_{\mu\upsilon}^k = q_{\mu\upsilon}^k + q_{\upsilon\mu}^k 
\end{equation}

Thridly, the three co-occurrence matrices $\mathcal{A}^k $ are linearly combined to get $\mathcal{A}'$ (Eq. 5)

\begin{equation}
	\mathcal{A}' = I + \alpha \cdot \mathcal{A}^1 + \beta \cdot \mathcal{A}^2 + \gamma \cdot \mathcal{A}^3 ,
\end{equation}
where $\alpha$, $\beta$ and $\gamma$ are hyperparameters.

Intuitively, the item pair will have closer relation if they are nearby in the sequence. Therefore, $\alpha$ is supposed to be the largest and $\gamma$ is supposed to be the smallest. The identity matrix $I$ is added into $\mathcal{A}'$ in order to facilitate the graph convolution.

Finally, inspired by GCN \cite{GCN}, LightGCN \cite{LightGCN} and GES \cite{GES}, a simple and lightweight graph convolution operation is used to obtain the global context embedding of each item (Eq. 6-7). The Laplace matrix $\mathcal{A}$ is first calculated by the obtained $\mathcal{A}'$ and its degree matrix $\mathcal{D}$ and then multiplied with the embedding lookup table of items $\mathbb{E}_{item}$ to get the global context embedding of items $\mathbb{E}_{global} \in \mathbb{R}^{|\mathcal{I}| \times d}$. 

\begin{equation}
	\mathcal{A} = \mathcal{D}^{-\frac{1}{2}} \mathcal{A}' \mathcal{D}^{-\frac{1}{2}} 
\end{equation}

\begin{equation}
	\mathbb{E}_{global} = \mathcal{A} \mathbb{E}_{item} 
\end{equation}

An example of the above calculations is illustrated in Figure 3. The GCE module only runs once in the training phase and can also be used as an independent module for any other sequential recommendation models.

\subsection{Recent Time Interval Representation Module}

After getting the global context embedding of items, our model further processes each user's recent item sequence and timestamp sequence to learn the user's personalized interaction preferences.

Firstly, the sequence length threshold $L_{rec}$ is set to intercept each user's most recent interactions, which does not include the target item for training. If there are less than $L_{rec}$ items, the padding strategy \cite{PIMI} will be used. Eventually, each user $u$ has a fixed-length sequence ${\mathcal{S}'}_u = \left\{i_1^u, i_2^u, \cdots ,i_{L_{rec}}^u\right\}$  and the corresponding timestamp sequence $\mathcal{{T}'}_u = \left\{t_1^u,t_2^u, \cdots ,t_{L_{rec}}^u\right\}$ .

Secondly, the time interval matrix is generated. This module resets $t_{\mu,\upsilon}^u$ as  $\min (|t_\upsilon^u - t_\mu^u |, L_{time})$ by using the threshold $L_{time}$ (defined in GCE) to emphasize the user’s recent interests. The interaction time interval of every two items in the user's recent interaction sequence $\mathcal{{S}'}_u$ is put into the time interval matrix $T^u \in \mathbb{R}^{L_{rec} \times L_{rec}}$, which is a symmetric matrix.

\begin{equation}
	T^u = {
		\begin{bmatrix}
			{t_{1,1}^u}&{t_{1,2}^u}&{\cdots}&{t_{1,L_{rec}}^u}\\
			{t_{2,1}^u}&{t_{2,2}^u}&{\cdots}&{t_{2,L_{rec}}^u}\\
			{\vdots}&{\vdots}&{\ddots}&{\vdots}\\
			{t_{L_{rec},1}^u}&{t_{L_{rec},2}^u}&{\cdots}&{t_{L_{rec},L_{rec}}^u}\\
	\end{bmatrix}}
\end{equation}

Thirdly, the time interval matrix $T^u$ is embedded to a tensor ${T'}^u \in \mathbb{R}^{L_{rec} \times L_{rec} \times d}$, where $d$ is the dimension of embedding. The time interval aware attention mechanism \cite{PIMI} is used to get attention scores $\mathbb{S}_1^u \in  \mathbb{R}^{L_{rec} \times L_{rec}} $ for each time interval in the sequence, denoted as

\begin{equation}
	\mathbb{S}_1^u = softmax({T'}^uW_1)^\top ,
\end{equation}
where $W_1$ is a trainable parameter matrix and $\top$ represents the transposition of the matrix.

Finally, the time interval embedding of items $\mathbb{E}_{time}^u$ in the recent sequence for each user $u$  is calculated through the broadcast mechanism in TensorFlow, which combines the attention score $\mathbb{S}^u$ and the time interval embedding tensor ${T'}^u$.

\begin{equation}
	\mathbb{E}_{time}^u = \mathbb{S}_1^u {T'}^u \in \mathbb{R}^{L_{rec} \times d}
\end{equation}

\subsection{Aggregation Module}

This module aggregates the global context embeddings and the time interval embeddings of the user's recent interactive items to get user’s personalized embedding.

Firstly, the items in each user's recent interaction sequence are picked out from the matrix $\mathbb{E}_{global}$ to form a global context embedding matrix $\mathbb{E}_{global}^u$ for the user $u$:

\begin{equation}
	\mathbb{E}_{global}^u = [\mathbf{e}_1^u, \mathbf{e}_2^u, \cdots, \mathbf{e}_{L_{rec}}^u] \in \mathbb{R}^{L_{rec} \times d}
\end{equation}

Secondly, the time interval embeddings and the global context embeddings of each user’s recent interactive items are combined by addition to get the hybrid embeddings:

\begin{equation}
	\mathbb{E}_{hyb}^u = \mathbb{E}_{time}^u + \mathbb{E}_{global}^u
\end{equation}

Thirdly, a GNN \cite{MTHEADATT} is used to further extract the user’s personalized representation from the recent sequence. To construct a graph, the adjacent items in the sequence are first connected and then a virtual central node $\mathbf{e}_\mathbf{c}^u \in \mathbb{R}^{1 \times d}$ is created and linked with all item nodes to capture the relation between items under larger than 1-hop. The embedding of $\mathbf{e}_\mathbf{c}^u$ is set to the average value of all item’s hybrid embeddings:

\begin{equation}
	\mathbf{e}_\mathbf{c}^{u(0)} = \frac{1}{L_{rec}} \sum_{i = 1}^{L_{rec}} \mathbb{E}_{hyb(i)}^u
\end{equation}

The multi-head attention mechanism is then used to aggregate the item nodes $\mathbb{E}_{hyb}^u$ and the central node $\mathbf{e}_\mathbf{c}^u$ in $L_{layer}$ steps. The layer-0 aggregation is the information initialization,

\begin{equation}
	q^{(0)}_i = \mathbb{E}_{hyb(i)}^u ,
\end{equation}

\begin{equation}
	k^{(0)}_i = concat[q^{(0)}_{i-1};\mathbf{e}_\mathbf{c}^{u{(0)}};q^{(0)}_{i};\mathbf{e}_i^u] ,
\end{equation}
where $i$ represents the $i$-$th$ item, and the superscript $(0)$ represents layer-0. The neighbor node $q^{(0)}_{i-1}$ , the center node $\mathbf{e}_\mathbf{c}^u$ and the global context representation $\mathbf{e}_i^u$ are concatenated. $q^{(0)}_i$ and $k^{(0)}_i$ are then sent into the multi-head attention mechanism at layer-$l$ $(l=1, \cdots, L_{layer})$,

\begin{equation}
	k^{(l)}_i = concat[q^{(l-1)}_{i-1}; \mathbf{e}_\mathbf{c}^{u(l-1)}; q^{(l-1)}_{i}; \mathbf{e}_i^u] ,
\end{equation}

\begin{equation}
	q^{(l)}_i = MultiHeadAtt(K=k^{(l)}_i, Q=q^{(l-1)}_i, V=k^{(l)}_i) .
\end{equation}

Similarly, the central node $\mathbf{e}_\mathbf{c}^u$ is aggretated as:
\begin{equation}
	k^{(l)} = concat[\mathbf{e}_\mathbf{c}^{u(l-1)};q^{(l)}_1; q^{(l)}_2; \cdots; q^{(l)}_{L_{rec}}] ,
\end{equation}

\begin{equation}
	\mathbf{e}_\mathbf{c}^{u(l)} = MultiHeadAtt(K=k^{(l)}, Q=\mathbf{e}_\mathbf{c}^{u(l-1)}, V=k^{(l)}) .
\end{equation}

After the total of $L_{layer}$ iterations, the final personalized embeddings of the items are stacked in sequential order into the matrix $\mathbb{E}^u $, which will be used in the next module.

\subsection{Multi-Interest Extraction Module}

This module extracts the $\mathcal{K}$ interests from the final representation of recent sequence.

Research has shown that the self-attention mechanism achieves better performance than capsule network in extracting user's multi-interests \cite{ComiRec}. Therefore, our model also uses the self-attention mechanism with two trainable parameter matrices $W_2 \in \mathbb{R}^{4d \times d}$ and $W_3 \in \mathbb{R}^{\mathcal{K} \times 4d}$ for user's $\mathcal{K}$ interest extraction:

\begin{equation}
	\mathbb{S}_2^u = softmax(W_3\tanh(W_2\mathbb{E}^{u\top}))
\end{equation}

\begin{equation}
	\mathbb{E}^u_{interest} = \mathbb{S}_2^u\mathbb{E}^u \in \mathbb{R}^{\mathcal{K} \times d}
\end{equation}

\subsection{Model Training}

To train the whole model, the optimal interest $o^u$ of the obtained $\mathcal{K}$ interests of each user is selected based on the target item (i.e., the next interactive ground-truth item) (Eq. 22),

\begin{equation}
	o^{u} = \mathbb{E}^u_{interest}[:, argmax(\mathbb{E}^{u\top}_{interest} \mathbf{e}_{target}^u)]
\end{equation}
where the $argmax$ operation in TensorFlow is used on the multiplication of $\mathbb{E}^u_{interest}$ and the embedding of the target item $\mathbf{e}_{target}^u \in \mathbb{E}_{global}$.

To be more efficient, the sampled softmax method \cite{SSL} is used to train the model, with the goal of minimizing the following loss $\mathcal{L}(\theta)$:

\begin{equation}
	\mathcal{L}(\theta) = \sum_{u\in\mathcal{U}} -\log \frac{\exp(o^{u\top}\mathbf{e}_{target}^u)}{\sum_{v \in Sampled(\mathcal{I})}\exp(o^{u\top}\mathbf{e}_v)}
\end{equation}

The optimizer used in GIMIRec is Adam \cite{Adam}.

\subsection{Recommendation Results}

After the training, GIMIRec achieves Top-$N$ recommendation for all users in the test set. Due to the large size of the item pool, a GPU-accelerated method Faiss \cite{Faiss} is used to search for $N$ items that are most relevant to user interests. The recommendation results are given based on the $\mathbb{E}^u_{ interest}$ obtained from Eq. 21,

\begin{equation}
	RecSet = GetTopN (\max \limits_{1\leq k \leq\mathcal{K}}(\mathbf{e}_{x}^{\top}\mathbb{E}^u_{interest(k)}))         
\end{equation}
where $\mathbf{e}_{x}$ is the embedding of the predicted item, $\mathbb{E}^u_{ interest (k)}$ is the $k$-$th$ interest embedding of the user $u$.

\section{Experiments}

This section introduces the experimental setup, evaluation methods and the performance comparison of the proposed GIMIRec model against six baselines, including the state-of-the-art multi-interest recommendation model. Moreover, the ablation study is carried out to verify the effectiveness of each innovation in GIMIRec.

\subsection{Experiment Settings}
{\bfseries Datasets.}  We conducted experiments on three real-world datasets: Amazon-Books \footnote{http://jmcauley.ucsd.edu/data/amazon/ \label{web}}, Amazon-Hybrid \textsuperscript{\ref {web}} and Taobao-Buy \footnote{https://tianchi.aliyun.com/dataset/dataDetail?dataId=649}. The Amazon dataset is divided into a series of sub-datasets according to the product category, among which Amazon-Books is the largest sub-dataset. Considering users are not only interested in books, we mixed five smaller datasets (`` Electronics '', `` Clothing, Shoes and Jewelry '', `` CDs and Vinyl '', `` Tools and Home Improvement '' and `` Grocery and Gourmet Food '') to form a new dataset Amazon-Hybrid. Taobao-Buy is the sub-dataset of the Taobao dataset filtering out all purchases. We discarded the users and items with less than 5 interactions and illegal timestamps. The length of the recent interaction sequence $L_{rec}$ is set to 20 for Amazon-Books and Amazon-Hybrid, and 50 for Taobao-Buy. The statistics of the datasets are listed in Table 1.

\begin{table}[h]
	\caption{Statistics of datasets}
	\label{tab:Statistics}
	\begin{tabular}{cccc}
		\toprule
		Dataset&users&items&interactions\\
		\midrule
		Amazon-Books&603,667&367,982&8,898,041\\
		Amazon-Hybrid&306,002&169,385&3,351,187\\
		Taobao-Buy&47,140&82,579&324,003\\
		\midrule
	\end{tabular}
\end{table}

We maintained the dataset division in the model validation of ComiRec \cite{ComiRec} and PIMI \cite{PIMI}. In detail, we divided each dataset into a training set, a validation set, and a test set at a ratio of 8:1:1. To evaluate the model, we predicted the most recent 20\% of user behaviors in the validation and test sets and used the remaining 80\% to infer user multi-interest embeddings.

{\bfseries Baselines.} Comparison methods include: (1) GRU4Rec \cite{GRU4Rec} (2015), the first sequential recommendation method using recurrent neural networks; (2) Youtube DNN \cite{YTBDNN} (2016), a deep neural network sequential recommendation model; (3) MIND \cite{MIND} (2019), the first sequential recommendation model proposing the multi-interest framework; (4) ComiRec-DR and (5) ComiRec-SA \cite{ComiRec} (2020) use dynamic routing mechanism and self-attention mechanism respectively to extract user’s multiple interests; (6) PIMI \cite{PIMI} (2021), the up-to-date sequential recommendation model based on multi-interest framework. It is noted that in above models based on multi-interest framework, MIND and ComiRec introduce extra information such as user age, item category etc., which is not adopted in PIMI and our model GIMIRec.

{\bfseries Evaluation Metrics.} We used three common Top-$N$ recommendation evaluation metrics, Hit Rate @$N$, Recall @$N$ and NDCG @$N$, to compare the performance of models. Hit Rate @$N$ represents the percentage that recommendation results contain at least one ground truth item in top $N$ position. Recall @$N$ indicates the proportion of the ground truth items in the recommendation results. NDCG (Normalized Discounted Cumulative Gain) @$N$ measures the ranking quality in the $N$ items by assigning high scores to high rankings \cite{PIMI}. In our experiments, $N$ is set to 20 and 50 respectively.

{\bfseries Implementation Details.} GIMIRec is implemented with Python 3.8 in TensorFlow 2.4, and a CUDA compatible GPU is used on the Ubuntu 20.04 operating system. The embedding dimension $d$ is set to 64 for all datasets. The batch size on Amazon-Book, Amazon-Hybrid and Taobao-Buy are set to 128,128 and 256, respectively, with a dropout rate 0.1 and learning rate 0.001. In sampled softmax loss calculation, the number of samples is set to 10. Other hyperparameter settings are shown in Table 2. Finally, we iterated the training up to 1 million rounds as the previous models did.

\begin{table}[h]
	\caption{Hyperparameter Settings}
	\label{tab:Setup}
	\begin{tabular}{cccccccc}
		\toprule
		Dataset&$a$&$b$&$\alpha$&$\beta$&$\gamma$&$\mathcal{K}$&$L_{time}$\\
		\midrule
		Amazon-Books&0.65&0.35&4.5&2&1&4&64\\
		Amazon-Hybrid&0.5&0.5&5&2.5&1&4&64\\
		Taobao-Buy&0.6&0.4&5&3&1&8&7\\
		\midrule
	\end{tabular}
\end{table}

\begin{table*}[h]
	\caption{Performance results of $N$=20 on three benchmark datasets (\%). The best in each column is bolded.}
	\label{tab:N20}
	\begin{tabular}{lccccccccc}
		\toprule
		\multirow{2}{*} &
		\multicolumn{3}{c}{Amazon-Books @20} & \multicolumn{3}{c}{Amazon-Hybrid @20} & \multicolumn{3}{c}{Taobao-Buy @20} \cr
		\cmidrule(lr){2-10} & Recall & NDCG & Hit Rate & Recall & NDCG & Hit Rate & Recall & NDCG & Hit Rate\cr
		\midrule
		GRU4Rec (2015)  & 4.003 & 6.792 & 8.815 & 3.178 & 2.850 & 6.598 & 0.887 & 0.660 & 1.633 \cr
		Youtube DNN (2016) & 4.364 & 7.679 & 10.585 & 3.395 & 3.012 & 7.173 & 1.279 & 1.118 & 2.375 \cr
		MIND (2019) & 4.652 & 7.943 & 10.638 & 4.315 & 3.938 & 8.585 & 3.802 & 3.126 & 6.235\cr
		ComiRec-DR (2020) & 5.344 & 9.125 & 12.001 & 2.722 & 2.437 & 5.657 & 1.265 & 1.096 &2.163 \cr
		ComiRec-SA (2020) & 5.602 & 8.993 & 11.488 & 3.817 & 2.992 & 7.555 & 7.802 & 6.662 & 11.538 \cr
		PIMI (2021) & 6.993 & 11.024 & 14.146 & 4.789 & 7.554 & 9.608 & 8.413 & 10.948 & 12.280 \cr
		\midrule
		{\bfseries GIMIRec} & \bfseries 9.180 & \bfseries 14.709 & \bfseries 18.149& \bfseries 7.033 & \bfseries 10.750 & \bfseries 13.304 & \bfseries 9.980 & \bfseries 13.051 & \bfseries 14.401 \cr
		{\bfseries Improvement \%} &  31.270&33.425&  28.298&  46.857 &  42.316 &  38.470 & 18.627 &  19.216 & 17.270 \cr
		\midrule
	\end{tabular}
\end{table*}

\begin{table*}[h]
	\caption{Performance results of $N$=50 on three benchmark datasets (\%). The best in each column is bolded.}
	\label{tab:N50}
	\begin{tabular}{lccccccccc}
		\toprule
		\multirow{2}{*} &
		\multicolumn{3}{c}{Amazon-Books @50} & \multicolumn{3}{c}{Amazon-Hybrid @50} & \multicolumn{3}{c}{Taobao-Buy @50} \cr
		\cmidrule(lr){2-10} & Recall & NDCG & Hit Rate & Recall & NDCG & Hit Rate & Recall & NDCG & Hit Rate\cr
		\midrule
		GRU4Rec (2015) & 6.541 & 10.319 & 13.546 & 5.564 & 3.820 & 11.441 & 1.864 & 1.076 & 3.097   \cr
		Youtube DNN (2016) & 7.410 & 12.025 & 15.906 & 5.767 & 3.781 & 11.660 & 2.190 & 1.396 & 3.754 \cr
		MIND (2019) & 7.731 & 12.032 & 16.128 & 6.807 & 4.826 & 13.375 & 5.186 & 3.338 & 8.038 \cr
		ComiRec-DR (2020) & 8.059 & 13.652 & 17.323 & 4.545 & 3.146 & 9.206 & 2.375 & 1.441 & 3.818 \cr
		ComiRec-SA (2020) & 8.469 & 13.567 & 17.127 & 6.697 & 4.154 & 12.941 & 9.570 & 7.322 & 14.168 \cr
		PIMI (2021) & 10.906 & 17.032 & 21.435 & 8.111 & 12.415 & 15.591 & 10.266 & 13.223 & 14.719 \cr
		\midrule
		{\bfseries GIMIRec} & \bfseries14.103 & \bfseries 21.756 & \bfseries 26.582 & \bfseries 11.642 & \bfseries 17.262 & \bfseries 21.186 & \bfseries 12.679 & \bfseries 16.424 & \bfseries 18.006 \cr
		{\bfseries Improvement \%}  &  28.986 &  27.271 &  22.959 &  43.537 &  39.039 & 35.884 &  23.515 &  24.211 &  22.334 \cr
		\midrule
	\end{tabular}
\end{table*}

\subsection{Comparisons and Analysis}

{\bfseries Recommendation performance.} The recommendation results of all methods on three datasets are shown in Tables 3 and 4, corresponding to the value of $N$=20 and $N$=50. To be fair, all models adopt the same parameter settings. The result shows that the GIMIRec model achieves the best performance on all evaluation indicators for all datasets. The performance of GIMIRec is significantly better than MIND and ComiRec even though there is no extra information introduced. Compared to the most advanced PIMI method, GIMIRec has great advantages, especially for Amazon data-sets. For Amazon-Hybrid, the performance improvement is the highest. The possible reason is that the multiple interests of users can be better represented from a rich variety of items. For Taobao-Buy, the performance improvement is not as good as that for Amazon datasets due to the shorter time span of dataset and sparser user interactions.

The result fully verifies the effectiveness of integrating the global context information.

\begin{table*}
	\caption{Ablation results of GIMIRec.}
	\label{tab:ExpAb}
	\resizebox{\linewidth}{!}{
		\begin{tabular}{lcccccccccccccccccc}
			\toprule
			\multirow{2}{*} &
			\multicolumn{3}{c}{Amazon-Books @20} & \multicolumn{3}{c}{Amazon-Hybrid @20} & \multicolumn{3}{c}{Taobao-Buy @20} &\multicolumn{3}{c}{Amazon-Books @50} & \multicolumn{3}{c}{Amazon-Hybrid @50} & \multicolumn{3}{c}{Taobao-Buy @50} \cr
			\cmidrule(lr){2-19} & Recall & NDCG & Hit Rate & Recall & NDCG & Hit Rate & Recall & NDCG & Hit Rate & Recall & NDCG & Hit Rate & Recall & NDCG & Hit Rate & Recall & NDCG & Hit Rate \cr
			\midrule
			GIMIRec$_{-INT}$ & 8.480&13.652&17.079&6.889&10.521&13.075&9.247&12.170&13.404&13.161&20.471&25.317&11.344&16.894&20.751&11.803&15.422&16.925\cr
			GIMIRec$_{-IN}$ & 8.267&13.269&16.940&6.748&10.313&12.794&9.843&12.884&14.380&13.201&20.467&25.638&11.282&16.745&20.499&12.173&15.587&17.264\cr
			GIMIRec$_{-I}$ & 8.852&14.195&17.632&6.566&10.036&12.346&9.366&12.224&13.616&13.646&21.056&25.870&10.709&15.977&19.405&12.016&15.404&17.031 \cr
			GIMIRec &\bfseries 9.180 & \bfseries 14.709 & \bfseries 18.149& \bfseries 7.033 & \bfseries 10.750 & \bfseries 13.304 & \bfseries 9.980 & \bfseries 13.051 & \bfseries 14.401 & \bfseries14.103 & \bfseries 21.756 & \bfseries 26.582 & \bfseries 11.642 & \bfseries 17.262 & \bfseries 21.186 & \bfseries 12.679 & \bfseries 16.424 & \bfseries 18.006 \cr
			\midrule
		\end{tabular}
	}
\end{table*}

\begin{table*}
	\caption{The impact of the number of interests $\mathcal{K}$ . The best in each column is bolded and the second best is underlined.}
	\label{tab:InterestNumCompare}
	\resizebox{\linewidth}{!}{
		\begin{tabular}{lcccccccccccccccccc}
			\toprule
			\multirow{2}{*} &
			\multicolumn{3}{c}{Amazon-Books @20} & \multicolumn{3}{c}{Amazon-Hybrid @20} & \multicolumn{3}{c}{Taobao-Buy @20} &\multicolumn{3}{c}{Amazon-Books @50} & \multicolumn{3}{c}{Amazon-Hybrid @50} & \multicolumn{3}{c}{Taobao-Buy @50} \cr
			\cmidrule(lr){2-19} & Recall & NDCG & Hit Rate & Recall & NDCG & Hit Rate & Recall & NDCG & Hit Rate & Recall & NDCG & Hit Rate & Recall & NDCG & Hit Rate & Recall & NDCG & Hit Rate \cr
			\midrule
			$\mathcal{K}=1$ & 8.635 &14.187& 17.934 & \underline{7.012}	&\underline{10.720}	&13.032& 7.853	&10.538 &	11.729 & 13.251&20.888&26.003&11.276&17.077&20.892&9.849&12.950&14.231\cr
			$\mathcal{K}=2$ &8.522	&13.872	&17.503 & 6.817	&10.512	&12.977	& 7.588	& 9.936	&11.092 &13.095&20.542&25.544&11.003&16.605&20.264&9.252&11.937&13.213\cr
			$\mathcal{K}=4$ & \bfseries 9.180	&\bfseries 14.709	& \bfseries 18.149	&\bfseries 7.033 &\bfseries 10.750	& \bfseries13.304&\bfseries 9.995 &	12.939&	14.295 &\bfseries 14.103&\bfseries 21.756&\bfseries 26.582&\bfseries 11.642&\bfseries 17.262&\bfseries 21.186& 12.168&15.907&17.540\cr
			$\mathcal{K}=6$ &8.721	&13.841	&17.008 & 6.557	& 10.080 &	12.343 & 9.855&	12.868	&14.295&13.516&20.809&25.403&10.720&16.015&19.441& 12.232&15.880&17.540\cr
			$\mathcal{K}=8$ & 8.739	&13.898	&17.086 & 7.007	& 10.632&\underline{13.222} & \underline{9.980}	& \bfseries 13.051&	\bfseries 14.401 &13.669&21.011&25.693&11.424&16.938&20.764& \bfseries 12.679&\bfseries 16.424&\bfseries 18.006\cr
			$\mathcal{K}=10$ &\underline{9.103}	&\underline{14.429}	&\underline{17.750} & 6.759	&10.365	&12.980& 9.856	&12.842	&\underline{14.316}&\underline{14.000}&\underline{21.570}&\underline{26.384}&11.326&16.871&20.774& \underline{12.423}&\underline{16.096}& \underline{17.773}\cr
			$\mathcal{K}=12$ &8.908	&14.100	&17.563 &6.991&	10.633&	13.209&9.913&\underline{12.959}	&\bfseries14.401&13.845&21.224&26.153&\underline{11.489}&\underline{17.122}&\underline{20.993}&12.387&15.910&17.625 \cr
			\midrule
		\end{tabular}
	}
\end{table*}

{\bfseries Ablation study.} We conduct a series of ablation studies to observe the impact of parameters settings in the GCE module on the results. GIMIRec is compared with three GIMIRec variants, GIMI-Rec$_{-INT}$ (no Interval, Number and Threshold), GIMIRec$_{-IN}$ (no Interval and Number) and GIMIRec$_{-I}$ (no Interval). Among them, GIMIRec$_{-I}$ changes GIMIRec by setting a=0 and b=1 in Eq. 2, indicating that the interaction interval information is not used when calculating the global context. GIMIRec$_{-IN}$ further changes GIMI-Rec$_{-I}$ in Eq. 3 by setting $q^k_{\mu\upsilon} = 1$, indicating that neither time intervals nor the co-occurrence number of item pairs are used. The difference of GIMIRec$_{-INT}$ and GIMIRec$_{-IN}$ is that the former does not set the time interval threshold $L_{time}$. The experimental results in Table 5 show that the integration of the number of interactions and time intervals in the global context extraction module in GIMIRec is effective and can further improve the performance. The use of the time interval threshold $L_{time}$ is valid to filter nearby item pairs.

{\bfseries Impact of the number of interests $\mathcal{K}$.} Different from the general sequential recommendation, the multi-interest framework learns $\mathcal{K}$ vectors for each user. Therefore, we discussed the impact of $\mathcal{K}$ on the performance. The experimental results are listed in Table 6. It shows that the recommendation result of modeling multiple vectors for user preferences is better than that of modeling only a single vector. Moreover, the optimal $\mathcal{K}$ value of the model relies on the dataset. The recommendation performance is the best when $\mathcal{K}$ is 4 for two Amazon datasets. However, for Taobao-Buy, most of the recommendation results are the best when $\mathcal{K}$ is 8. The possible reason is that Taobao-Buy has more product categories, which could reflect more interests of users.

{\bfseries Impact of hyperparameters $\alpha$, $\beta$ and $\gamma$.} In the GCE module, we used three hyperparameters, $\alpha$ , $\beta$ and $\gamma$ , to set the co-occurrence weight of 1-hop, 2-hop and 3-hop respectively. Table 7 lists the recommendation results of GIMIRec on Amazon-Hybrid under different combinations of three parameters. It shows that the result with all three matrices is better than that of only 1-hop matrix. The recommendation result is optimal when the proportion of $\alpha$ , $\beta$ and $\gamma$ is 5:2.5:1. The result is consistent with expectation, that is, the 1-hop relation is the most important.

\begin{table}
	\caption{Impact of hyperparameters $\alpha$, $\beta$ and $\gamma$. The best in each column is bolded and the second best is underlined.}
	\label{tab:ExpNHops}
	\resizebox{\columnwidth}{!}{
		\begin{tabular}{ccccccc}
			\toprule
			\multirow{2}{*} &
			\multicolumn{3}{c}{Amazon-Hybrid @20} & \multicolumn{3}{c}{Amazon-Hybrid @50} \cr
			\cmidrule(lr){2-7} & Recall & NDCG & Hit Rate & Recall & NDCG & Hit Rate \cr
			\midrule
			1:0:0 &6.248&9.599&11.905&10.183&15.251&18.617\cr
			1:1:0 &6.916&10.603&13.189&11.291&16.803&20.637 \cr
			1:1:1 &6.872&10.498&13.016&11.323&16.864&20.679 \cr
			3:2:1 & 6.550&10.014&12.326&10.614&15.844&19.290\cr
			4:2:1 &\underline{6.986}&\underline{10.679}&\underline{13.251}&\underline{11.420}&\underline{16.972}&\underline{20.839} \cr
			4.5:2:1 &6.906&10.590&13.117&11.326&16.877&20.735\cr
			4:2.5:1 & 6.503&9.997&12.294&10.705&15.953&19.349\cr
			5:2.5:1 &\bfseries7.033&\bfseries 10.750&\bfseries 13.304&\bfseries 11.642&\bfseries 17.262&\bfseries 21.186 \cr
			5:3:1 & 6.729&10.239&12.686&11.144&16.530&20.124\cr
			\midrule
		\end{tabular}
	}
\end{table}

{\bfseries Impact of hyperparameters $a$ and $b$.} The $a$ and $b$ in Eq. 2 determine how the GCE module uses the time interval of historical interactions. The value of $a$ (the weight of time interval) has been proved to be effective in above ablation experiments. In Table 8, we experimented different combinations of $a$ and $b$ in Amazon-Hybrid. As seen, the best setting is $a$=0.5, $b$=0.5 and a too large value of $a$ would cause the deterioration in model performance, because it may weaken the impact of co-occurrence number. 

\begin{table}
	\caption{The impact of hyperparameters $a$ and $b$.}
	\resizebox{\columnwidth}{!}{
		\begin{tabular}{lcccccc}
			\toprule
			\multirow{2}{*} &
			\multicolumn{3}{c}{Amazon-Hybrid @20} & \multicolumn{3}{c}{Amazon-Hybrid @50} \cr
			\cmidrule(lr){2-7} & Recall & NDCG & Hit Rate & Recall & NDCG & Hit Rate \cr
			\midrule
			$a$=0.3, $b$=0.7 & 7.009&10.657&13.232&11.489&17.045&20.836  \cr
			$a$=0.4, $b$=0.6 & 6.696&10.228&12.670&11.075&16.550&20.254 \cr
			$a$=0.5, $b$=0.5 &\bfseries7.033&\bfseries 10.750&\bfseries 13.304&\bfseries 11.642&\bfseries 17.262&\bfseries 21.186  \cr
			$a$=0.6, $b$=0.4 &7.022&10.677&13.248&11.534&17.139&20.986\cr
			$a$=0.7, $b$=0.3 &6.545&10.072&12.382&10.758&16.073&19.545\cr
			\midrule
		\end{tabular}
	}
\end{table}

{\bfseries Impact of the time interval threshold $L_{time}$. } Time thresholds are set in both the GCE module and the recent time interval representation module. We take Amazon-Hybrid as an example to observe the impact of different $L_{time}$ on the results (see Table 9). It shows that for Amazon-Hybrid the model with $L_{time}$ set to 64 performs best, probably because a larger time interval threshold may cause user’s recent interests to not be well emphasized, while a smaller threshold may cause the loss of part of user’s interests. We also found that the optimal $L_{time}$ in Taobao-Buy is 7 (see the settings in Table 2). The results demonstrate that the $L_{time}$ value is closely related to the dataset.

\begin{table}
	\caption{The impact of the time threshold $L_{time}$.}
	\label{tab:ExpTimeThreshold}
	\resizebox{\columnwidth}{!}{
		\begin{tabular}{lcccccc}
			\toprule
			\multirow{2}{*} &
			\multicolumn{3}{c}{Amazon-Hybrid @20} & \multicolumn{3}{c}{Amazon-Hybrid @50} \cr
			\cmidrule(lr){2-7} & Recall & NDCG & Hit Rate & Recall & NDCG & Hit Rate \cr
			\midrule
			$L_{time}$ =32 &6.984&10.601&13.225&11.379&16.911&20.784\cr
			$L_{time}$ =64 &\bfseries7.033&\bfseries 10.750&\bfseries 13.304&\bfseries 11.642&\bfseries 17.262&\bfseries 21.186 \cr
			$L_{time}$ =128 &6.977&10.590&13.026&11.331&16.934&20.702\cr
			$L_{time}$ =256 &6.856&10.470&13.039&11.307&16.812&20.660\cr
			\midrule
		\end{tabular}
	}
\end{table}

{\bfseries Impact of the number of GNN layers.} The aggregation module uses a multi-head attention mechanism for aggregation and the number of GNN layers directly affects the performance. We tried 1 to 5 layers of aggregation on the Amazon-Hybrid dataset and found that the performance is stabilized after 3 layers. The detailed results are shown in Figure 4.

\begin{figure}[h]
	\centering
	\includegraphics[width=\linewidth]{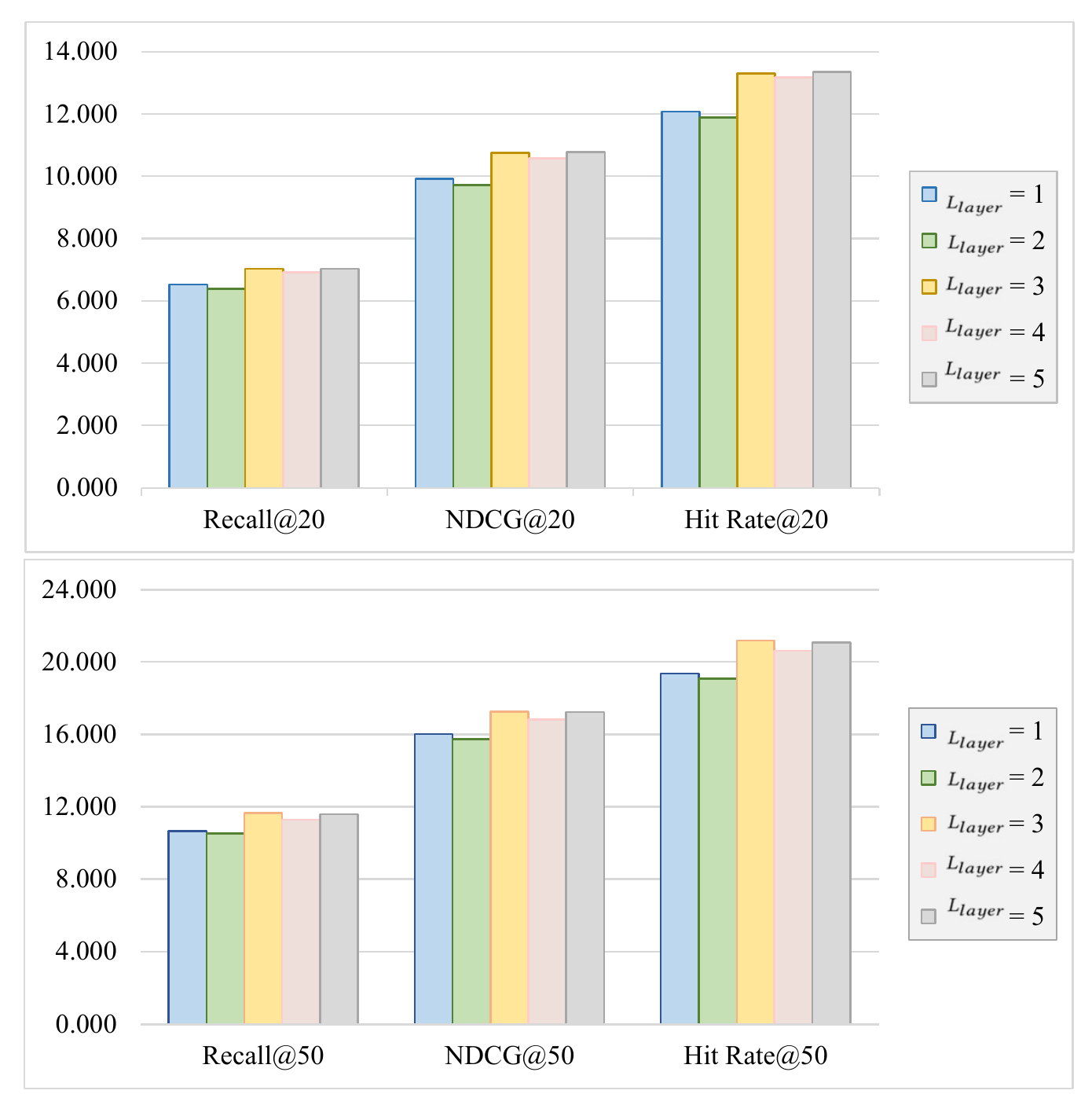}
	\caption{The impact of the number of GNN layers $L_{layer}$. }
	\Description{Compare result.}
\end{figure}

{\bfseries Time complexity analysis of the GCE module.} In GCE, we conducted a lightweight graph convolution instead of a complete GCN. The overall time complexity of the latter is $\mathcal{O}(|\xi|dK + |\mathcal{I}|d^2K) $ \cite {GCN1, GCN2}, where $|\mathcal{I}|$ represents the number of items, $|\xi|$ represents the number of interactions (edges), $d$ is the embedding dimension and $K$ is the depth (i.e., the number of layers) of GCN. However, in the GCE module, only 1, 2 and 3-hop relations are calculated, the process of graph convolution is simplified and only one convolution layer is used. As a result, the time complexity of GCE is reduced to $\mathcal{O}(|\xi|d)$.

\section{Conclusion}
This paper proposes a novel multi-interest framework for sequential recommendation, which leverages the historical interactions of all users to extract global context information, so as to better learn users’ multiple interests and provide better personalized recommendations. Specifically, the co-occurrence number of item pairs and their time intervals from all interaction sequences are first used to learn the global context embedding of items based on a simplified graph convolution operation. Then, the global context embedding is fused with the recent time interval embedding to get the personalized embedding for each user. Finally, a multi-interest framework is used to learn $\mathcal{K}$ interest vectors for each user to provide recommendations. Extensive experiments verify that the potential information contained in the historical sequences can greatly benefit sequential recommendation. Since the proposed module does not use any external information and has lightweight calculations, it can be easily transplanted to any sequential recommendation model. In the future, we will further explore the multi-interest framework by focusing on the non-linear relationship between multiple interests of users and fine-grained representation of items.

\bibliographystyle{ACM-Reference-Format}
\bibliography{sample-base}

\end{document}